\documentclass{ws-ijmpe}

\begin{document}

\markboth{P. Magierski}{}

%%%%%%%%%%%%%%%%%%%%% Publisher's Area please ignore %%%%%%%%%%%%%%%
%
\catchline{}{}{}{}{}
%
%%%%%%%%%%%%%%%%%%%%%%%%%%%%%%%%%%%%%%%%%%%%%%%%%%%%%%%%%%%%%%%%%%%%

\title{IN-MEDIUM ION MASS RENORMALIZATION AND LATTICE VIBRATIONS
       IN THE NEUTRON STAR CRUST}

\author{\footnotesize PIOTR MAGIERSKI}

\address{Faculty of Physics, Warsaw University of Technology,\\
ul. Koszykowa 75, 00-662 Warsaw, Poland \\
Piotr.Magierski@olimp.if.pw.edu.pl}

\maketitle

\begin{history}
\received{October 13, 2003}
\revised{October 27, 2003}
%\accepted{(Day Month Year)}
%\comby{(xxxxxxxxxx)}
\end{history}

\begin{abstract}
The inner crust of a neutron star consists of nuclei immersed in a
superfluid neutron liquid. As these nuclei move through the
fermionic medium they bring it into motion as well. As a result their
mass is strongly renormalized and  the spectrum of the ion
lattice vibrations is significantly affected. Consequently
we predict that the specific heat and the
lattice thermal energy of the Coulomb crystal at these densities
are noticeably modified.
\end{abstract}

In the standard picture the inner crust of a neutron star
consists of neutrons, protons and electrons.
At low temperatures ( $T\lesssim 1$ MeV) one expects the presence of atomic nuclei (ions)
immersed in a neutron gas. At the densities:
$10^{11}$ g/cm$^{3}\lesssim\rho\lesssim 10^{14}$ g/cm$^{3}$ they form a crystal
lattice stabilized by the Coulomb interaction.
The electrons which are ultrarelativistic at these densities form a
strongly degenerate, uniform gas.
The detailed structure of this part of the star has been the subject
of a considerable theoretical effort
(.\cite{bbp,nvb,pra,dhm,dha,mh,mbh} and references therein).
At the bottom of the crust the appearance of non-spherical nuclei,
forming exotic structures, has been predicted. In this region
besides the Coulomb interaction also the quantum effects, associated with
a neutron scattering on nuclear inhomogeneities, play an important role,
leading eventually to a disordered phase .\cite{bma1,mh}.

In order to understand various processes associated with
thermal evolution of a neutron star, the properties of
Coulomb crystals in the inner crust have been studied by many authors
(see e.g. .\cite{bpy} and references therein).
The plasma frequency of the system reads:
$\omega_{p} = \sqrt{4\pi \rho_{ion} Z^{2}e^{2}/M}$,
where $\rho_{ion}$ denotes the ion density, $Z$ is the proton number of an ion,
and $M$ is the ion mass. Since $\rho_{ion}=1/\frac{4}{3}\pi R_{c}^{3}$, where
$R_{c}$ is the Wigner-Seitz cell radius then:
\begin{equation}
\omega_{p} = \sqrt{\frac{3 Z^{2}e^{2}}{R_{c}^{3} M}}.
\end{equation}
The quantities: $Z$ and $R_{c}$, in the above equation, can be
obtained from calculations based on a density functional
method. However, the determination of the ion mass $M$
requires a certain caution, since
the nucleus is immersed in the fermionic environment and its bare mass has
to be renormalized to take into account the interaction with surrounding
neutrons. If the neutrons outside a nucleus
have formed a normal Fermi gas the problem
would have been rather complicated.\cite{rosch}.
In the inner crust, however, it simplifies due to the fact
that at low temperatures, $T\lesssim 0.1$ MeV, the neutron
gas is predicted to be strongly paired. In this case one may apply
the hydrodynamic approximation which describe the system
by an irrotational velocity field.
Although at finite temperatures there always exists a gas of quasiparticle
excitations which form a normal flow, at temperatures much lower
than the critical temperature its influence can be neglected.

We assume in the following that the nucleus has
a uniform nucleon density $\rho_{in}$ and fills a sphere of radius $R$.
The neutron gas is characterized by the density $\rho_{out}$.
We assume moreover that the flow velocity is below
the critical velocity for loss of superfluidity.

In order to calculate the mass $M$ of an ion, the Poisson
equation for the velocity field has to be solved, both inside and
outside a nucleus. The boundary conditions read:
\begin{eqnarray}
\Phi_{in}|_{r=R} &=& \Phi_{out}|_{r=R}, \\
\rho_{in}(\frac{\partial}{\partial r}\Phi_{in} - \vec{n}\cdot\vec{u})|_{r=R}&=&
\rho_{out}(\frac{\partial}{\partial r}\Phi_{out} - \vec{n}\cdot\vec{u})|_{r=R}, \\
\Phi_{out}|_{r\rightarrow\infty} &=& 0,
\end{eqnarray}
where $\Phi_{in}$ and $\Phi_{out}$ stand for the velocity field inside and
outside a nucleus, respectively. The vector $\vec{n}$ is an outward normal
to the surface of a spherical nucleus and $r$ is a radial coordinate (the
center of a nucleus is placed at $r=0$). The velocity of the surface element
was denoted by $\vec{u}$. The first equation stems from
the requirement that the phase of the wave function of the superfluid system
is continuous, the second one is a conservation of mass
in radial flow, and finally the third one ensures that the correct asymptotic behavior
for the velocity field is obtained. The knowledge of the velocity
field allows us to determine the kinetic energy and corresponding effective ion
mass. Consequently, the following expression for the mass of an
ion being a subject of a translational motion can be derived:
\begin{equation}
M^{ren}=\frac{4}{3}\pi R^{3} m \rho_{in} \frac{(1-\gamma)^{2}}{2\gamma + 1},
\end{equation}
where $\gamma=\rho_{out}/\rho_{in}$, and $m$ is a nucleon mass.

The above expression relates the mass of an ion to the density
of the neutron gas. Note, that for a uniform system, i.e.
$\gamma = 1$ the ion mass is equal to zero.
Since the nucleon density vary substantially across
the inner crust, one expects that the properties of the Coulomb crystal
will be noticeably modified. Indeed, the plasma frequency
in the inner crust changes as compared to the value calculated
for the case: $\rho_{out}=0$ (ion in the vacuum). Namely,
$\displaystyle{\frac{\omega_{p}}{\omega^{ren}_{p}}}=
\displaystyle{\frac{|1-\gamma|}{\sqrt{2\gamma + 1}}}$, where
$\omega^{ren}_{p}$ corresponds to the plasma frequency for the system
with renormalized ion masses.

\begin{figure}[th]
\centerline{\psfig{file=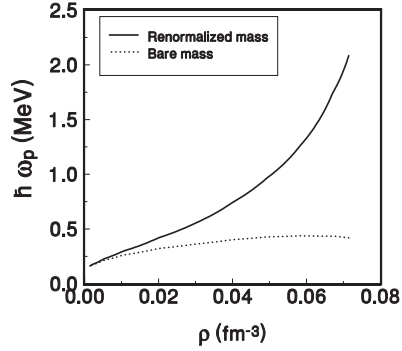,width=7.5cm}}
\vspace*{4pt}
\caption{Plasma frequency in the inner crust
as a function of the nucleon density.
Nuclear parameters were taken from .\protect\cite{dhm,hzd}.}
\end{figure}
\begin{figure}[th]
\centerline{\psfig{file=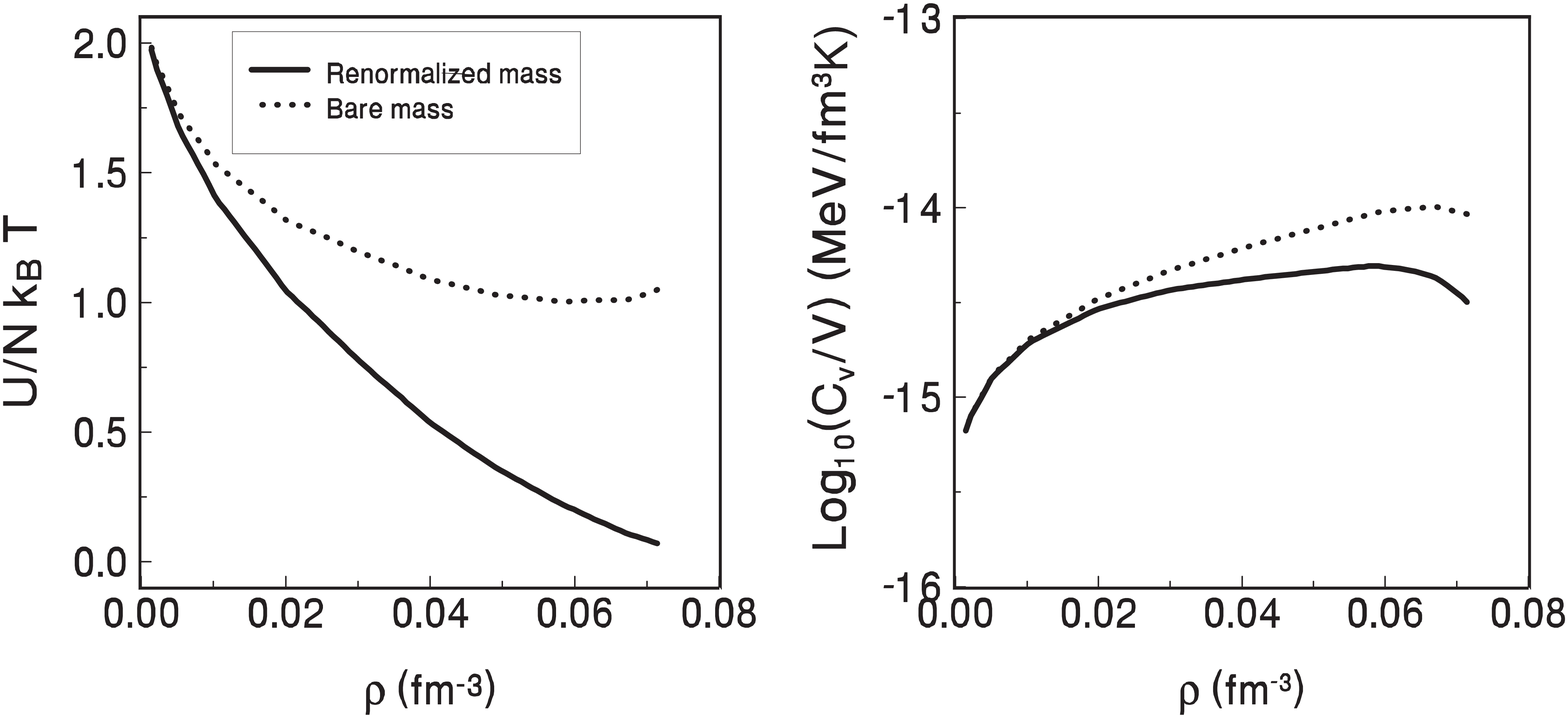,width=14cm}}
\vspace*{4pt}
\caption{The reduced lattice thermal energy: $u = U/N k_{B} T$ (left figure) and
specific heat per unit volume (right figure) at the temperature $T=0.1$ MeV,
as a function of the nucleon density for a bcc Coulomb crystal.
$N$ is the number of ions and $k_{B}$ denotes the Boltzmann constant.}
\end{figure}

In the Fig. 1 we have plotted $\hbar\omega_{p}$
and $\hbar\omega^{ren}_{p}$ as a function of the nucleon density.
One can notice large discrepancies between the value obtained using the
bare nuclear mass and the renormalized one. It clearly indicates that
the Coulomb crystal becomes more stiff at the bottom of the crust
\footnote{One expects the phase transition to non-spherical nuclear
phases to occur at the bottom of the crust. We disregard this effect in this paper.}.
Consequently the density of the phonon spectrum is also modified
as compared to the case of bare ion masses.
The thermodynamic functions: the lattice thermal energy and the specific heat
have been shown in the Fig. 2.
One can notice that the renormalized values at a given density
and temperature are smaller. In the limit
of $T\rightarrow 0$ the analytic expressions for the ratios of
lattice thermal energy $U$ and specific heat $C_{v}$ can be derived.
Applying the phenomenological relations from the ref.\cite{bpy}  one gets:
\begin{equation}
\frac{U^{ren}(T=0)}{U(T=0)}=
\frac{C_v^{ren}(T=0)}{C_{v}(T=0)}=\frac{|1-\gamma|^{3}}{(2\gamma+1)^{3/2}},
\end{equation}
where $U^{ren}$ and $C_{v}^{ren}$ denote
the renormalized values of lattice energy and specific heat, respectively.

The following conclusions can be drawn from the presented results:
\begin{enumerate}
\item There is a substantial renormalization effect of a nuclear/ion mass
in the inner crust of a neutron star, due to the presence of a superfluid
neutron liquid.
\item Consequently, the phonon spectrum and thermodynamic
functions of the Coulomb crystal are significantly altered.
\item Thermal and electric conductivities of the inner crust, governed
by the electron-phonon scattering are expected to be modified. In particular,
the contributions coming from Umklapp processes have to be recalculated using
the renormalized ion masses.
\item There is clearly a need for consideration of other nuclear degrees of freedom
like e.g. shape vibrations, which may significantly modify the electron
transport properties.
\end{enumerate}

\section*{Acknowledgements}
This research was supported in part by the Polish Committee
for Scientific Research (KBN) under Contract No.~5~P03B~014~21.
Numerical calculations were performed
at the Interdisciplinary Centre for Mathematical and Computational
Modelling (ICM) at Warsaw University. Author would like to thank
Aurel Bulgac for discussions and valuable comments.

\end{document}